\documentclass[prl,showpacs,twocolumn]{revtex4-1} 
\usepackage{latexsym}
\usepackage{mathrsfs}
\usepackage{amssymb}
\usepackage{amsmath}
\usepackage{braket}
\usepackage[usenames,dvipsnames]{xcolor}

\newcommand{\A}{\mathbb{A}}
\newcommand{\C}{\mathbb{C}}

\newcommand{\CP}{\mathbb{CP}}

\newcommand{\PA}{\mathbb{PA}}
\newcommand{\PT}{\mathbb{PT}}

\renewcommand{\P}{\mathbb{P}}

\newcommand{\cH}{\mathcal{H}}

\newcommand{\T}{\mathbb{T}}

\newcommand{\p}{\partial}
\newcommand{\dbar}{\bar\partial}
\newcommand{\e}{\mathrm{e}}

\newcommand{\cA}{\mathcal{A}}
\newcommand{\cV}{\mathcal{V}}

\newcommand{\cM}{\mathcal{M}}
\newcommand{\cN}{\mathcal{N}}

\renewcommand{\P}{\mathbb{P}}

\newcommand{\vol}{\mathrm{Vol}}

\newcommand{\GL}{\mathrm{GL}}

\newcommand{\rd}{\, \mathrm{d}}

\newcommand{\1}{{\rm 1\hskip-0.25em I}}
\newcommand{\be}{\begin{equation}\label}
\newcommand{\ee}{\end{equation}}
\newcommand{\bea}{\begin{eqnarray}\label}
\newcommand{\eea}{\end{eqnarray}}

\newcommand{\la}{\langle}
\newcommand{\ra}{\rangle}

\begin{document}

\title{Ambitwistor strings in 4-dimensions}
\author{Yvonne Geyer, Arthur E. Lipstein, Lionel Mason}
\affiliation{Mathematical Institute, Andrew Wiles Building,\\Woodstock Road, Oxford, OX2 6CG, UK}

\begin{abstract}
We develop ambitwistor string theories for 4 dimensions to obtain new formulae for tree-level gauge and gravity amplitudes with arbitrary amounts of supersymmetry.  Ambitwistor space is the space of complex null geodesics in complexified Minkowski space, and in contrast to earlier ambitwistor strings,  we use twistors rather than vectors to represent this space.  Although superficially similar to the original twistor string theories of Witten, Berkovits and Skinner, these theories differ in the assignment of worldsheet spins of the fields, rely on both twistor and dual twistor representatives for the vertex operators, and use the ambitwistor procedure for calculating correlation functions. Our models are much more flexible, no longer requiring maximal supersymmetry, and the resulting formulae for amplitudes are simpler, having substantially reduced moduli. These are supported on the solutions to the scattering equations refined according to MHV degree and can be checked by comparison with corresponding formulae of Witten and of Cachazo and Skinner. 
\end{abstract}

\maketitle

\section{Introduction}
The twistor string theories of Witten, Berkovits and Skinner\cite{Witten:2003nn,Berkovits:2004hg, Skinner:2013xp} not only led to remarkable new formulae for tree-level scattering amplitudes, but also provided a tantalising paradigm for how twistor theory might eventually make contact with physics.  However, it is still a long way from being fully realized both because of the reliance on maximal supersymmetry of these models and the lack of a clear route to an extension to a critical model in which loop calculations will make sense.  More recently it has become clear that there are many such remarkable tree-level formulae for scattering amplitudes \cite{Witten:2004cp,Hodges:2012ym,Cachazo:2012kg,Cachazo:2012pz, Cachazo:2013gna,Cachazo:2013hca, Cachazo:2013iea}.  One family \cite{Cachazo:2013hca} were seen to arise naturally  from string theories in ambitwistor space, the space of complex null geodesics \cite{Mason:2013sva}.  Such ambitwistor spaces can be defined in all dimensions and the string is critical in 10 dimensions.  It provides an infinite tension chiral limit of the conventional RNS superstring.  An important advantage over the original twistor-strings is that it extends naturally to provide a tentative candidate for the full field theory all-loop integrand \cite{Adamo:2013tsa} (albeit one that is likely to reproduce standard field theory divergences). 

Ambitwistor strings can be defined almost algorithmically by complexifying the action for a spinning massless particle.  While the focus of \cite{Mason:2013sva} was the RNS model,
we specialize to four spacetime dimensions in this paper. Ambitwistor space then has an alternative spinorial representation in which the constraints $P^2=0$ are explicitly solved. The resulting ambitwistor string models arise  as the complexifications of the four-dimensional Ferber superparticle \cite{Ferber:1977qx}.  Indeed the original twistor-string  was similarly interpreted in \cite{Siegel:2004dj,Berkovits:2005} and the similarity with the ambitwistor approach was also remarked upon in \cite{Mason:2013sva,Bandos:2014lja}.  Using the spinorial representation as the target space, we construct ambitwistor string models for Yang-Mills theory and gravity in four dimensions with any amount of supersymmetry. These models yield remarkably simple new formulae for the tree-level scattering amplitudes which are parity invariant, supported on the scattering equations, and dependent on very few moduli. There is also a reasonably clear route to an extension to a critical theory that will be valid at loops by reduction from a critical model such as \cite{Berkovits:2013xba} 

\section{Ambitwistor strings in 4 dimensions}
Projective ambitwistor space $\P\A$ is a supersymmetric extension of the space of
complex null geodesics. In four dimensions, ambitwistor space can be expressed as a quadric $Z\cdot W=0$ inside $\PT\times \PT^*$.  Here we work with $\cN$  supersymmetries so that $Z=(\lambda_\alpha,\mu^{\dot\alpha},\chi^r)\in \T=\C^{4|\cN}$,  $W=(\tilde \mu ,\tilde \lambda, \tilde \chi)\in\T^*$ with $\chi, \tilde \chi$ fermionic, $\alpha=0,1$, $\dot\alpha=\dot 0,\dot 1$ chiral spinor indices and $r=1,\ldots \cN$ R-symmetry indices.  Ambitwistor space $\A$ is the set $Z\cdot W=0$ where
$$
Z\cdot W:=\lambda_\alpha\tilde\mu^\alpha+\mu^{\dot\alpha}\tilde\lambda_{\dot\alpha}+\chi^r\tilde\chi_r\, ,
$$
where we also quotient by the scalings $\Upsilon-\tilde\Upsilon$ where $\Upsilon= Z\cdot \p/\p Z$ and $\tilde \Upsilon=W\cdot\p/\p W$.  We also have the incidence relations
\begin{eqnarray}
\mu^{\dot \alpha}&=&i(x^{\alpha\dot\alpha} +i\theta^{r\alpha}\tilde\theta^{\dot\alpha}_r)\lambda_\alpha\, ,\qquad  \chi^r=\theta^{r\alpha}\lambda_\alpha\, , \nonumber \\
\tilde \mu^{ \alpha}&=&-i(x^{\alpha\dot\alpha} -i\theta^{r\alpha}\tilde\theta^{\dot\alpha}_r)\tilde \lambda_{\dot \alpha}\, ,\qquad
\tilde \chi_r=\tilde \theta_r^{\dot \alpha}\tilde \lambda_{\dot \alpha}\, ,
\end{eqnarray}
which realize a point $(x,\theta,\tilde\theta)$ in (non-chiral) super Minkowski space as a quadric,  $\CP^1\times\CP^1$ parametrized by $(\lambda,\tilde\lambda)$.  It is easily seen that these lie inside the set $Z\cdot W=0$ and indeed, these are the only quadrics in $\PA$ of that degree. To make contact with null geodesics, the momenta can be defined to be $P_{\alpha\dot\alpha}=\lambda_\alpha\tilde\lambda_{\dot\alpha}$, which now automatically satisfy the constraint $P^2=0$.  The symplectic potential is $\Theta =\frac i2(Z\cdot \rd W -  W\cdot \rd Z) $. 

Our ambitwistor string consists of worldsheet fields  $(Z,W)$ that are spinors on the worldsheet Riemann surface $\Sigma$ and take values in $\T\times \T^*$. The action is based on the symplectic potential and the constraint $Z\cdot W=0$ imposed by Lagrange multiplier $a$, a $(0,1)$-form on $\Sigma$, so that 
$$
S=\frac1{2\pi}\int_\Sigma  W\cdot \dbar Z-Z\cdot \dbar W  +a Z\cdot W\, .
$$
 There is a gauge symmetry associated with $a$ that quotients by $\Upsilon - \widetilde \Upsilon$.  A key difference between this theory and the original Berkovits-Witten theory is that we  fix $(Z,W)$ to be spinor fields on the worldsheet.  


After adding worldsheet gravity 
by starting with the operator $\bar\p+e\p$, we gauge fix $e=0$ leading to a ghost $(b,c)$ system.  Similarly we gauge fix $a$ to obtain the natural $\dbar$-operator on the spin bundle  and introduce associated ghosts $(u,v)$.  We end up with a BRST operator 
$$
Q=\int c T + uZ\cdot W\, 
$$
where $T$ is the world-sheet stress tensor. This will be anomalous  (i.e., $Q^2\neq0$) in general, although there should be choices of matter that give an anomaly free theory.  At tree level however, this won't be relevant, and  the ghost system will serve to give the $GL(2,\C)$ quotients that are needed in the tree-level formulae.
 Amplitudes will be obtained as correlation functions of vertex operators.  In the following we will just give  integrated vertex operators (they simply differ by a factor of $c$ from their unintegrated counterparts) and will instead just divide by the volume of $\GL(2,\C)$ in the final formula, understood in the usual Faddeev-Popov sense.  

\section{Yang-Mills amplitudes} 
Yang-Mills vertex operators arise from general wave functions  $\alpha$ that are $\dbar$-closed $(0,1)$-forms multiplied by the currents $J\cdot t_a$ of some current-algebra $J$ associated to some Lie algebra, of which the $t_a $ are elements (hereon $a=1,\ldots ,n$  indexes particle number), to give $\cV_a=\int_\Sigma \alpha J\cdot t_a$.  In general, such an $\alpha$ corresponds to an off-shell Maxwell field on space-time, but if it extends off $\PA$ into $\PT\times \PT^*$ to 3rd order or beyond, it must be on-shell (see for example \cite{Baston:1987av}), and only when on-shell is it manifestly $Q$-closed. On shell, such wave functions are a sum of wave-functions pulled back from either twistor space or dual twistor space, thus leading to two different types of vertex operators.  For momentum eigenstates, 
\begin{eqnarray}
\cV_a'&=&\int\frac{\rd s_{a}}{s_{a}}\bar{\delta}^{2}(\lambda_{a}-s_{a}\lambda)\e^{is_{a}\left([\mu\,\tilde{\lambda}_{a}]+\chi^{r}\tilde{\eta}_{ar}\right)}J\cdot t_{a} \, \\
\widetilde  \cV_a&=& \int\frac{\rd s_{a}}{s_{a}}\bar{\delta}^{2}(\tilde{\lambda}_{a}-s_{a}\tilde{\lambda})\e^{is_{a}\left(\la\tilde{\mu}\,\lambda_{a}\ra+\tilde{\chi}_{r}\eta_{a}^{r}\right)}J\cdot t_{a}\, 
\end{eqnarray}
where for a complex variable $z$, $\bar\delta(z)= \dbar( 1/2\pi iz)$.  These can be seen to be straighforwardly $Q$-invariant.  However,
having the supersymmetry in this form will be inconvenient in
what follows. A more convenient representation is obtained by a Fourier transform of the $\tilde
\eta$s into $\eta$s in the first type of vertex operator,
\be{}
\cV_a=\int \frac{\rd s_a}{s_a} \bar\delta^{2|\cN}(\lambda_a-
s_a\lambda|\eta_a-s_a\chi)\e^{is_a[\mu\,
  \tilde\lambda_a]}J\cdot t_a \, .
\ee
where for a fermionic variable $\chi$, $\delta(\chi)=\chi$.  We obtain full $\mathcal{N}=4$ Yang-Mills amplitudes with the above vertex operators for $\cN=3$ (with $r=1,\ldots,\cN=4$ we would have double the spectrum).

N$^{k-2}$MHV Yang-Mills amplitudes will be obtained as correlation functions of the above vertex operators taking $k$ from dual twistor space and $n-k$ from twistor space:
$$
\cA=\left\la \widetilde \cV_1 \ldots \widetilde \cV_k \cV_{k+1}\ldots  \cV_n\right\ra\, .
$$
The current algebra correlator gives the Parke-Taylor denominator (together with some multitrace terms that we will ignore for the purposes of this paper).  As in \cite{Mason:2013sva}, rather than attempt to compute the infinite number of contractions required by the exponentials,  we instead take the exponentials into the action to provide sources 
\[
\int_\Sigma  \sum_{i=1}^k  i s_i (\la \tilde \mu \lambda_i\ra +\tilde \chi\cdot \eta_i)\bar\delta(\sigma-\sigma_i) + \sum_{p=k+1}^nis_p[\mu\,\tilde\lambda_p]\bar\delta(\sigma-\sigma_p). 
\] 
The equations of motion for $Z$ and $W$ are then
\begin{eqnarray}
\bar{\partial}_\sigma Z&=&\bar{\partial}\left(\lambda,\mu,\chi\right)=\sum_{i=1}^{k}s_{i}\left(\lambda_{i},0,\eta_{i}\right){\bar\delta}\left(\sigma-\sigma_{i}\right),  \nonumber \\ \label{dbar-l}
\bar{\partial}_\sigma W&=&\bar{\partial}\left(\tilde{\mu},\tilde{\lambda},\tilde{\chi}\right)=\sum_{p=k+1}^{n}s_{p}\left(0,\tilde{\lambda}_{p},0\right){\bar\delta}(\sigma-\sigma_{p}) .
\end{eqnarray}
Since $(Z,W)$ are worldsheet spinors, the solutions are uniquely given by
\begin{eqnarray}\nonumber
Z(\sigma)&=&\left(\lambda,\mu,\chi\right)=\sum_{i=1}^{k}\frac{s_{i}\left(\lambda_{i},0,\eta_{i}\right)}{\sigma-\sigma_{i}}
\\
W(\sigma)&=&\left(\tilde{\mu},\tilde{\lambda},\tilde{\chi}\right)=\sum_{p=k+1}^{n}\frac{s_{p} \left(0,\tilde{\lambda}_{p},0\right)}{\sigma-\sigma_{p}} \, .
\end{eqnarray}

With this, we are left with the integrals
\begin{multline}\label{final-form}
\cA=\int  \frac1{\vol \,\GL(2,\C)} \prod_{a=1}^n\frac {\rd
  s_a\rd\sigma_a}{s_a (\sigma_a-\sigma_{ a+1})}  
  \prod_{i=1}^k  \bar\delta^2(\tilde \lambda_i -s_i\tilde\lambda)\\ \prod_{p=k+1}^n\bar\delta^{2|\cN} (\lambda_p-s_p\lambda(\sigma_p),\eta_p-s_p\chi(\sigma_p))  .
\end{multline}
We can write this in terms of homogeneous coordinates on the Riemann sphere $\sigma_\alpha=\frac1s (1,\sigma)$ using the notation $(i \,j)=\sigma_{i\alpha}\sigma_j^\alpha$ (with indices raised and lowered by the usual skew symmetric $\epsilon_{\alpha\beta}$) as follows
\be{dbar-sol-hgs}
Z(\sigma)= \sum_{i=1}^{k}\frac{\left(\lambda_{i},0,\eta_{i}\right)}{(\sigma\, \sigma_{i})}
\, , \qquad 
W(\sigma)= \sum_{p=k+1}^{n}\frac{(0,\tilde{\lambda}_{p},0) }{(\sigma \,\sigma_{p})}\, 
\ee
(where we have rescaled $W$ and $Z$ by a factor of $1/s$) and
\begin{multline}\label{final-form-hgs}
\cA=\int  \frac1{\vol \,\GL(2,\C)} \prod_{a=1}^n\frac {\rd^2 \sigma_a}{(a\, a+1)}  \,
\prod_{i=1}^k  \bar\delta^2(\tilde \lambda_i - \tilde\lambda(\sigma_i))  \\ \prod_{p=k+1}^n\bar\delta^{2|\cN} (\lambda_p-\lambda(\sigma_p),\eta_p-\chi(\sigma_p))\, .
\end{multline}
For notational simplicity we have taken the colour order to be $(1,\ldots,n )$; of course any other choice will just lead to the obvious re-ordering of the Parke-Taylor denominator.
There are $2n$ bosonic delta functions and  $2n-4$ integrals, the minus four coming from the $\vol (\GL(2,\C))$ quotient, yielding four momentum-conserving delta functions. Momentum conservation can be seen from
\[
\sum_{p=k+1}^{n}\lambda_{p}\tilde{\lambda}_{p}=\sum_{p=k+1}^{n}\tilde{\lambda}_{p}\sum_{j=1}^{k}\frac{\lambda_{j}}{(p \, j)}=-\sum_{j=1}^{k}\lambda_{j}\tilde{\lambda}_{j},
\]
where we used the first (second) set of delta functions in \eqref{final-form-hgs} to get the first (second) equality. Similarly $\sum_{a=1}^{n}\tilde{\lambda}_{a}\eta_{a}=0$.

Defining $P(\sigma)=\lambda(\sigma)\tilde\lambda(\sigma)$, the scattering equations $\lambda_a^\alpha\tilde\lambda_a^{\dot\alpha}\cdot P_{\alpha\dot\alpha}(\sigma_a)=0$ also follow on the support of the delta functions.  Indeed, these are here refined to give just those appropriate to N$^k$MHV degree as 
$$
[\tilde \lambda_i\, \tilde\lambda(\sigma_i)]=0\, , \; i=1\ldots k\,,\quad \la \lambda_p \, \lambda(\sigma_p)\ra=0 , \; p=k+1 \ldots n.
$$

The formula \eqref{final-form-hgs} can be verified at $\cN=0$ by comparison with  Witten's formula 3.22  in \cite{Witten:2004cp} (or analagous formulae in \cite{Cachazo:2013gna}) and extended to arbitrary $\cN$ by superconformal invariance.
This comparison follows by integrating out $2n$ of the $4n$ moduli in Witten's formula against $2n$ of the delta functions leaving just those for the $2n$ homogeneous coordinates $u_a$. This determines $\lambda(u)$ and $\tilde \lambda(u)$ and leads directly to our formula after identifying  $u_i= \sigma_i/\prod_{j=k+1}^n(\sigma_j\, \sigma_i)$ for $i=1,\ldots,k$ (see \cite{tbp} for more details).

This model will also have vertex operators leading to amplitudes of a nonminimal conformal gravity like that of Berkovits and Witten for the original twistor string.


\section{Einstein Gravity Amplitudes}
For Einstein gravity we construct an ambitwistor analogue of David Skinner's model \cite{Skinner:2013xp} as follows.  This model has  fields $(Z,W)$ that are worldsheet spinors with values in $\T\times \T^*$ as before, and
$(\rho,\tilde\rho)$ again in $\T\times \T^*$ but with opposite statistics (i.e.\ taking values in $\C^{\cN| 4}$ rather than $\C^{4|\cN}$).  In order to break conformal invariance we introduce
infinity twistors $I_{IJ}$ and $I^{IJ}$ that in general can encode a cosmological constant and a gauging of $R$-symmetry, but are rank 2 in the simplest zero cosmological constant ungauged case that we will work with here setting $I_{IJ}Z_1^IZ_2^J=\la \lambda_1\, \lambda_2\ra=:\la Z_1, Z_2\ra$ and $I^{IJ}W_{1I}W_{2J}=[\tilde\lambda_1\, \tilde\lambda_2]=:[W_1, W_2]$.   
We furthermore gauge the following  currents
$$
K_a=\left(Z\cdot W, \rho\cdot \tilde \rho,  W\cdot \rho, [W, \tilde \rho], Z\cdot \tilde \rho, \la Z, \rho\ra, \la \rho, \rho\ra, [\tilde \rho, \tilde \rho]\right)\,,
$$ 
which leads to the introduction of the corresponding weight zero ghosts $(\beta_a,\gamma^a)$, 
together with the fermionic $(b,c)$ ghosts as before \cite{Adamo:2013tca}.  These lead to a
BRST $Q$-operator
\begin{equation}
Q=\int c T + \gamma^aK_a -\frac i2 \beta_a \gamma^b\gamma^c  C^a_{bc}\, ,
\end{equation}
where $C^a_{bc}$ are the structure constants of the current algebra $K_a$. 

In this Einstein gravity model,   $Q$-invariance implies that vertex operators are built from $\dbar$-closed $(0,1)$-forms $h$ of weight two on twistor space and similarly $\tilde h$ from dual twistor space.  
For momentum eigenstates, $h$ and $\tilde{h}$ are given by
\begin{align}
 h_a&=\int \frac{\rd s_a}{s_a^3}\bar\delta^{2|\cN}(\lambda_a-
 s_a\lambda|\eta_a-s_a\chi)\e^{is_a[\mu\,
  \tilde\lambda_a]}\\
 \tilde{h}_a&=\int\frac{\rd s_{a}}{s_{a}^{3}}\bar{\delta}^{2}(\tilde{\lambda}_{a}-s_{a}\tilde{\lambda})\e^{is_{a}\left(\la\tilde{\mu}\,\lambda_{a}\ra+\tilde{\chi}_{r}\eta_{a}^{r}\right)}\,.
\end{align}
These yield two types of vertex operators, appearing in integrated or unintegrated form, here integration being with repect to ghost zero modes.   The ghosts $\gamma=(\gamma^3,\gamma^4)$, $\nu=(\gamma^5,\gamma^6)$ each have one zero mode and these can be fixed by the insertion of one each of the unintegrated vertex operators
$$
 V_h=\int_{\Sigma}\delta^2(\gamma)h\, , \qquad  \widetilde V_{\tilde{h}}=\int_{\Sigma}\delta^2(\nu)\tilde{h}\,.
$$
As usual, the remaining states are represented by integrated vertex operators
\begin{align}
 \mathcal{V}_h&=\int \Big[W,\frac{\p h}{\p Z}\Big]+\Big[\tilde{\rho},\frac{\p}{\p Z}\Big]\,\rho\cdot\frac{\p h}{\p Z},\\
 \widetilde{\mathcal{V}}_{\tilde{h}}&=\int\Big\la Z,\frac{\p \tilde{h}}{\p W}\Big\ra+\Big\la\rho,\frac{\p}{\p W}\Big\ra\tilde{\rho}\cdot\frac{\p \tilde{h}}{\p W}.
\end{align}
For full $\cN=8$ supergravity  we can again use the above vertex operators for $\cN=7$. 
This suggests an interesting connection with Hodges' $\cN=7$ formalism, \cite{Hodges:2011wm}.

Amplitudes are now given by the worldsheet correlation function
\begin{equation}
 \cM=\left\langle \widetilde V_{\tilde h_1}\prod_{i=2}^{k} \widetilde{\mathcal{V}}_{\tilde h_i} \prod_{p=k+1}^{n-1} \mathcal{V}_{{h}_p} V_{{h}_n}\right\rangle.
\end{equation}
A correlator of  a fermion system, here  $(\rho,\tilde \rho)$, is the determinant of a matrix of possible contractions.  Here we extend this to the following $n\times n$ matrix:
\begin{equation*}
\cH= \begin{pmatrix}{ \mathbb{H}}& 0\\ 0&\widetilde{\mathbb{H}}\end{pmatrix},
\end{equation*}
where, for $i,j\in\{1,...,k\}$ and $p,q\in\{k+1,...,n\}$
\begin{align}
 &{\mathbb{H}}_{ij}=\frac{\braket{i\, j}}{(i\, j)}, &i\neq j, \qquad & {\mathbb{H}}_{ii}=-\sum_{j=1,j \neq i}^{k} {\mathbb{H}}_{ij}\\
 &\widetilde{\mathbb{H}}_{pq}=\frac{[p\, q]}{(p\, q)}, & p\neq q, \qquad & \widetilde{\mathbb{H}}_{pp}=-\sum_{q=k+1,q \neq p}^n \widetilde{\mathbb{H}}_{pq}.
\end{align}
The off-diagonal element $\cH_{ij}$ is the contraction of the $\rho$-term in the $i$th vertex operator with the $\tilde \rho$-term in the $j$th, and the diagonal elements of $\cH$ come from the remaining terms in the integrated vertex operators. Repeating the steps for Yang-Mills, we obtain for  gravity amplitudes 
\begin{multline} \label{Mgrav}
 \cM=\int \frac{\prod_{a=1}^n \rd^2 \sigma_a}{\vol \,\GL(2,\C)}\: \text{det}'(\cH)\, \prod_{i=1}^k  \bar\delta^2(\tilde \lambda_i - \tilde\lambda(\sigma_i))\\
 \prod_{p=k+1}^n\bar\delta^{2|\cN} (\lambda_p-\lambda(\sigma_p),\eta_p-\chi(\sigma_p))\, ,
\end{multline}
where $\det' \cH$ is the determinant omitting a row and column from each of $\widetilde{\mathbb H} $ and $\mathbb{H}$ corresponding to the unintegrated vertex operators;  the answer is independent of this choice because each has kernel $(1,\ldots,1)$.

The equivalence to the formula of \cite{Cachazo:2012kg} is seen by following the Yang-Mills strategy and making judicious identifications of reference spinors with the given $\sigma_a$ \cite{tbp}.

\section{Conclusion}
Ambitwistor strings provide a chiral infinite tension limit of conventional strings.  Here we have formulated them in four space-time dimensions in terms of twistors and dual twistors, leading to remarkably simple new formulae for tree amplitudes for (super) Yang-Mills and (super) gravity. These are nontrivially related to previous twistor string formulae, as we describe in detail in \cite{tbp}. Our gravitational formula is similar to the link representation of \cite{He:2012er}, and so one can regard ambitwistor strings as providing the theory underlying such representations.


There are many directions for future exploration. One important question regards the representation of loop amplitudes. Although our model is sufficient for computing tree-level amplitudes, in general it is noncritical and anomalous (the gauge anomalies require $\cN=4$ for the first model and $\cN=8$ for the Einstein gravity model, which suggest a doubling of the spectrum in our context). On the other hand, it is likely that a critical, anomaly-free theory can be obtained by coupling to appropriate matter as for example obtained by reduction from an anomaly-free theory in 10 dimensions \cite{Mason:2013sva,Adamo:2013tsa, Berkovits:2013xba}. It might then be possible to represent loop amplitudes as integrals over higher genus moduli spaces of maps. 

An issue raised by the gauging associated with $a$ is the validity of our imposition of the choice of degree of the line bundles on $\Sigma$  in which $(Z,W)$ take their values.    Often one would sum over the degrees of the line bundle spanned by $\Upsilon - \widetilde \Upsilon$ as will be discussed in more detail in \cite{tbp} where it will be seen that different choices give the same answer or zero so the answer is only changed by an overall constant.

Another direction is the generalization of our formulae to nonzero
cosmological constant; our model already allows for this, 
leading to non-zero entries in the off-diagonal blocks of $\cH$, and could provide an efficient method for computing tree-level correlation functions in $AdS_{4}$  and $dS_{4}$. These can be compared to the formulae of \cite{Adamo:2012nn,Adamo:2013tja} and may in turn have applications to the $AdS_{4}/CFT_{3}$  correspondence and cosmology. 


{\it Acknowledgements:}
We would like to thank Andrew Hodges and David Skinner for helpful discussions. AL is supported by a
Simons Postdoctoral Fellowship, YG by the EPSRC and the Mathematical Prizes fund and LM is supported by EPSRC grant number EP/J019518/1. 

%

\end{document}